\documentclass{article}

\usepackage[english]{babel}

\usepackage[letterpaper,top=2cm,bottom=2cm,left=3cm,right=3cm,marginparwidth=1.75cm]{geometry}

\usepackage{amsmath}
\usepackage{graphicx}
\usepackage[colorlinks=true, allcolors=blue]{hyperref}

\title{Integrating AI's Carbon Footprint into Risk Management Frameworks: Strategies and Tools for Sustainable Compliance in Banking Sector}
\author{Nataliya Tkachenko (Ph.D., MA)\footnote{This work has been conducted as part of the Author's Visiting Fellowship at the Centre for Finance, Technology \& Regulation (Judge Business School, University of Cambridge), and we are grateful for the opportunity to have this work hosted by the School.}}

\begin{document}
\maketitle

\begin{abstract}
This paper examines the integration of AI's carbon footprint into the risk management frameworks (RMFs) of the banking sector, emphasising its importance in aligning with sustainability goals and regulatory requirements. As AI becomes increasingly central to banking operations, its energy-intensive processes contribute significantly to carbon emissions, posing environmental, regulatory, and reputational risks. Regulatory frameworks such as the EU AI Act, Corporate Sustainability Reporting Directive (CSRD), Corporate Sustainability Due Diligence Directive (CSDDD), and the Prudential Regulation Authority’s SS1/23 are driving banks to incorporate environmental considerations into their AI model governance. Recent advancements in AI research, like the Open Mixture-of-Experts (OLMoE) framework and the Agentic RAG framework, offer more efficient and dynamic AI models, reducing their carbon footprint without compromising performance. Using these technological examples, the paper outlines a structured approach for banks to identify, assess, and mitigate AI's carbon footprint within their RMFs, including adopting energy-efficient models, utilising green cloud computing, and implementing lifecycle management. The paper also emphasises cross-departmental collaboration within banks and the use of respective enabling tools such as the GHG Protocol Toolkit for carbon accounting and Fairlearn for AI fairness assessment. Building on the experts’ opinions, such as UNEP FI, we argue that targeting alignment with the global standards and directives like IFRS, ESRS, and GFANZ, banks can enhance their sustainability reporting and model governance, ensuring long-term resilience. By effectively integrating these practices into their RMFs, banks can navigate the regulatory landscape, mitigate environmental risks, and promote a sustainable future in banking, balancing innovation with environmental responsibility.

\end{abstract}

\section{Introduction}

The banking sector is increasingly integrating artificial intelligence (AI) into its operations to improve efficiency, customer service, risk management, and overall operational effectiveness \cite{s1,s2}. AI systems power critical banking functions, from customer service chatbots and credit scoring algorithms to fraud detection and automated financial advice. However, the growing reliance on AI brings with it a significant environmental cost \cite{s3}. The training, deployment, and ongoing operation of AI systems consume large amounts of energy, contributing to a substantial carbon footprint. In the context of the banking sector, this environmental impact presents a new dimension of risk that must be managed alongside traditional financial and operational risks.

To address these questions, recent AI research has branched into three additional areas of responsible AI, notably: ways of enhancing computational efficiency, complementarity of language models with smarter frameworks, and development of agentic models for more dynamic workflows. These advances have the potential to mitigate the carbon footprint of AI by making models more efficient and adaptable. From the regulatory perspective, UK Prudential Regulation Authority’s SS1/23 \cite{s4, s5} has emphasised the need for banks to enhance model governance and oversight, indirectly influencing the sustainability of AI models by promoting efficient model management practices. The European Union has taken a proactive stance in addressing the environmental and ethical concerns of AI through regulations such as the Corporate Sustainability Reporting Directive (CSRD), the Corporate Sustainability Due Diligence Directive (CSDDD), and the EU AI Act. 

The CSRD mandates comprehensive disclosure of sustainability metrics, including the environmental impacts of corporate activities, while the CSDDD aims to strengthen corporate accountability for sustainability across the EU, requiring companies to conduct due diligence on their environmental and social impacts. Additionally, the EU AI Act emphasises the ethical deployment of AI, including transparency, accountability, and risk management \cite{s6, s7, s8}. For banks, these regulations, along with the expectations of SS1/23, mean that the carbon footprint of their AI systems must be recognized, measured, and incorporated into their risk management frameworks (RMFs). Failure to do so could result in regulatory penalties, reputational damage, and a misalignment with stakeholder expectations around environmental responsibility. This essay explores how banks can effectively incorporate the carbon footprint of AI into their RMFs, ensuring compliance with regulatory requirements while promoting sustainable AI practices.

\section{Understanding the Carbon Footprint of AI}

AI's carbon footprint stems primarily from its energy-intensive processes \cite{s2}. Large-scale AI models, especially those using deep learning techniques, require significant computational power for training and deployment. Training complex models involves processing vast amounts of data, often using high-performance hardware like graphics processing units (GPUs) in data centres. This process can consume more energy than entire countries in some cases. For instance, training a single deep learning model can produce as much carbon as five cars over their lifetimes \cite{s9}. In the banking sector, where AI models must often be retrained and updated to remain accurate and compliant with evolving regulations like SS1/23, this energy consumption can be a continuous source of carbon emissions.

The operational deployment of AI systems in customer banking, such as real-time fraud detection, credit scoring, and automated customer service, further contributes to the carbon footprint. These systems require ongoing data processing and computational resources, resulting in a persistent energy demand. Additionally, the infrastructure supporting these AI applications, including data storage and cloud services, adds to the overall energy consumption. When scaled across an entire banking institution, the cumulative carbon footprint of AI can become significant, offsetting the potential efficiency gains that these systems provide \cite{s10, s13}.

Recent research in AI has sought to address this issue by focusing on enhancing computational efficiency. For instance, the Open Mixture-of-Experts (OLMoE)\footnote{OLMoE: \url{https://github.com/allenai/OLMoE/blob/main/README.md}} framework, developed by the Allen Institute of AI, uses a sparse mixture-of-experts approach, activating only a subset of parameters during model inference. This technique achieves up to sevenfold efficiency gains over dense models while maintaining or even exceeding performance. Similarly, OneGen\footnote{OneGen: \url{https://github.com/zjunlp/OneGen}} introduces a unified architecture for generation and retrieval tasks, reducing inference time and costs while improving task performance. Another advancement, Feature Sampling and Partial Alignment Distillation (FSPAD), accelerates language model inference by up to 13.3\% and enables smaller models like LLaMA 3.1 to outperform larger models like GPT-4. These innovations reflect a growing trend towards more efficient AI models, which is crucial for reducing the carbon footprint associated with AI operations in banking and finance \cite{s11, s12, s14}.

Recognizing the environmental impact of AI is critical not only for regulatory compliance but also for aligning with broader corporate sustainability goals. We therefore put forward the proposition that the carbon footprint of AI must be considered alongside the positive contributions of AI to efficiency and customer service to achieve a holistic view of the banks' environmental impacts. Incorporating AI's carbon footprint into the RMFs of banks is thus a strategic imperative to address environmental risks, comply with regulatory standards, and maintain the bank's reputation as a responsible corporate citizen \cite{s15}.

\section{Incorporating AI's Carbon Footprint into RMFs}

Building on the well-documented best practices, it can be argued that incorporating the carbon footprint of AI into the risk management frameworks of banks can involve several steps, from risk identification and assessment to the development and implementation of mitigation strategies. This process must be integrated with the bank's existing risk management practices to ensure a cohesive approach to managing all dimensions of risk, including environmental, operational, and regulatory risks \cite{s16, s18}.

The first step in including AI’s carbon footprint in the RMF is risk identification \cite{s17}. This usually involves recognising the potential environmental risks associated with the deployment and operation of AI systems. Direct risks include the high energy consumption associated with training and running AI models, which contributes to carbon emissions and climate change. The more complex and data-intensive the AI model, the greater its energy consumption and carbon footprint. In the context of banking, AI systems used for credit scoring, fraud detection, and customer service are likely to be among the most energy-intensive due to the need for real-time processing and high accuracy. Additionally, SS1/23's emphasis on model governance requires banks to identify the environmental risks associated with their models, prompting them to include considerations of energy efficiency and carbon emissions as part of their overall risk management strategy \cite{s4, s5}.

In addition to direct environmental risks, banks must also consider regulatory risks. With the introduction of regulations such as the CSRD, the CSDDD, and SS1/23, banks face the risk of non-compliance if they fail to adequately measure, disclose, and mitigate the environmental impact of their AI systems \cite{s18}. The CSDDD sets the stage for stronger corporate accountability and sustainability in the EU, requiring companies to conduct due diligence on the environmental and social impacts of their operations. The latest CDP Policy Explainer \cite{s27} provides a detailed roadmap for banks to address these CSDDD requirements and align them with CDP disclosures, including the carbon footprint of AI. The guide covers climate transition plans in alignment with global standards such as IFRS S2, ESRS, SEC, GRI, and GFANZ, highlighting clear transition plan elements like governance, scenario analysis, risk management, strategy, financial planning, and target setting. SS1/23 complements this by urging banks to incorporate environmental metrics into their model risk assessments, ensuring that banks evaluate their AI models not only for performance and robustness but also for their sustainability impact \cite{s19, s20, s21, s22}.

By integrating AI's carbon footprint into their RMFs, banks can align their practices with these standards and frameworks \cite{s23}. The CDP Policy Explainer emphasises actionable insights across governance to value chain engagement, providing a clear path for banks to report on their climate risks, opportunities, and progress. This not only ensures compliance with CSDDD requirements but also helps banks stay ahead of the regulatory curve by demonstrating a commitment to sustainability and responsible AI use.

Once the risks have been identified, the next step is to assess and measure the carbon footprint of AI within the RMF. This involves quantifying the emissions associated with AI models throughout their lifecycle, from development and training to deployment and maintenance. To facilitate this process, banks can leverage carbon accounting tools such as the GHG Protocol Toolkit and OpenLCA \cite{s24, s25}. These tools enable banks to calculate the greenhouse gas (GHG) emissions generated by their AI systems, providing a concrete measure of the environmental impact. Metrics such as energy consumption per transaction, carbon emissions per AI model, and the total carbon footprint of AI-related activities can be used as key performance indicators (KPIs) for monitoring the environmental impact. In line with SS1/23, banks must also include these environmental metrics in their model risk assessment processes, treating carbon emissions as a specific risk category alongside traditional risk metrics.

In addition to using carbon accounting tools, banks can conduct scenario analysis to understand how different AI deployment strategies affect their overall carbon footprint \cite{s23}. For instance, banks can evaluate the impact of using more energy-efficient AI models or shifting AI processing to cloud providers that use renewable energy sources. Innovations such as OLMoE, OneGen, and FSPAD present opportunities for banks to reduce their carbon footprint by implementing more computationally efficient models. By analysing various scenarios, banks can also identify strategies that minimise the carbon footprint while still achieving operational and business objectives. The CSDDD, CDP frameworks, and SS1/23 all offer guidance on how to incorporate scenario analysis into transition plans, emphasising the importance of assessing climate-related risks and opportunities within the bank's value chain \cite{s20, s21, s22, s23, s24, s25}.

With the carbon footprint assessed and measured, banks can then implement risk mitigation strategies to reduce the environmental impact of their AI systems. One effective strategy is the adoption of energy-efficient AI models and algorithms. Techniques such as model pruning, quantisation, and knowledge distillation can reduce the computational requirements of AI models without compromising their performance. For example, model pruning involves removing unnecessary parameters from a neural network, resulting in a smaller and more energy-efficient model. Similarly, quantisation reduces the precision of the numbers used in a model's calculations, leading to reduced energy consumption. The use of advanced frameworks, such as MemoRAG\footnote{MemoRAG: \url{https://github.com/qhjqhj00/MemoRAG}} and GraphInsight\footnote{GraphInsight: \url{https://github.com/CarloNicolini/GraphInsight}}, further optimises AI performance by complementing language models with smarter frameworks, enhancing efficiency and reducing energy demands. Under SS1/23's model lifecycle management principles, banks are encouraged to continuously monitor these models to ensure ongoing compliance with both performance and sustainability standards \cite{s26, s29, s30}.

Banks can also mitigate the carbon footprint of AI by adopting green cloud computing practices. This involves using cloud service providers that operate data centres powered by renewable energy sources. By offloading AI training and processing to green cloud providers, banks can significantly reduce the carbon emissions associated with their AI operations. Additionally, banks can optimise the energy efficiency of their AI systems by implementing lifecycle management practices as emphasised in SS1/23. This includes regularly auditing AI models to ensure they are operating efficiently, decommissioning outdated models, and using AI systems judiciously to avoid unnecessary energy consumption. SS1/23’s lifecycle management focus provides banks with a framework to assess the carbon impact of AI models from development to decommissioning, making it easier to identify opportunities for reducing emissions \cite{s30, s31, s32, s33}.

The final step in incorporating AI’s carbon footprint into the RMF is integration with the bank's existing risk management processes. To effectively manage the environmental risks associated with AI, banks should include AI's carbon footprint as a specific category within their RMF. This category should be treated with the same level of importance as traditional risk categories such as credit, market, and operational risks. Regular monitoring and reporting of the carbon footprint of AI should be incorporated into the bank’s risk monitoring processes. By adapting tools like Carbon Tracker\footnote{Carbon Tracker Tools: \url{https://carbontracker.org/engagement-tools/}} and TransitionArc\footnote{TransitionArc: \url{https://climatearc.org/news/introducing-transitionarc}}, banks can continuously monitor the energy consumption and carbon emissions of their AI systems, allowing them to take proactive measures if emissions exceed predefined thresholds. SS1/23’s emphasis on model monitoring and oversight supports this approach by ensuring that banks have robust processes in place to track and manage the environmental performance of their AI models throughout their lifecycle \cite{s34, s35, s36}.

Furthermore, we recommend that banks define their risk appetite and tolerance levels for AI’s carbon footprint \cite{s31, s32, s33}. This involves setting thresholds for acceptable levels of carbon emissions associated with AI operations and establishing escalation procedures if these thresholds are breached \cite{s34, s35, s36}. By clearly defining the bank's risk tolerance, risk managers can make informed decisions about AI deployment strategies that balance operational efficiency with environmental responsibility. For such cases, potential suggested integration of advanced agentic models, such as the Agentic RAG framework\footnote{Agentic RAG framework: \url{https://lilianweng.github.io/posts/2023-06-23-agent/}} for time series analysis, could provide dynamic workflows for reducing the carbon footprint of AI \cite{s13}. By coordinating sub-agents for tasks like forecasting and anomaly detection, this framework reduces mean absolute error (MAE) and mean absolute percentage error (MAPE), enhancing efficiency and reducing the energy consumption of AI models used in customer banking. In alignment with SS1/23, banks can leverage these innovative models to ensure that the performance and sustainability of their AI systems remain balanced, whilst aligned with regulatory expectations.

\section{Aligning with Regulatory Requirements}

Aligning the incorporation of AI's carbon footprint in RMFs with regulatory requirements is critical for ensuring compliance and avoiding legal and reputational risks. The CSRD requires banks to disclose the environmental impacts of their operations, including the use of AI systems. By including the carbon footprint of AI in their RMFs, banks can ensure that they meet CSRD's sustainability disclosure requirements. This involves reporting on key environmental metrics such as energy consumption, carbon emissions, and the measures taken to reduce the carbon footprint of AI systems. These disclosures should be integrated into the bank's broader sustainability reporting, providing stakeholders with a transparent view of the bank's environmental performance \cite{s31}.

The CSDDD further strengthens corporate accountability by requiring companies to conduct due diligence on the environmental and social impacts of their operations. By including the carbon footprint of AI in their RMFs, banks can conduct the necessary due diligence to identify, prevent, and mitigate environmental risks associated with AI deployment. Mentioned before in this paper, the CDP Policy Explainer provides a roadmap for aligning with the CSDDD and global standards, including IFRS, ESRS, and GFANZ. By following this roadmap, banks can ensure that their climate transition plans address critical elements such as governance, scenario analysis, risk management, strategy, financial planning, and target setting \cite{s8}.

In addition to CSRD and CSDDD compliance, the EU AI Act \cite{s1} imposes specific obligations on the deployment of AI systems, particularly those classified as high-risk. The Act requires banks to ensure that their AI systems are used ethically and sustainably, with a focus on transparency and accountability. Banks must include carbon footprint assessments in their AI audits to comply with the Act's provisions. This involves documenting the environmental impact of AI systems, particularly those used in critical functions such as credit scoring and fraud detection. By conducting regular AI audits that include carbon footprint assessments, banks can demonstrate their commitment to ethical AI use and sustainable practices. SS1/23 complements these regulatory requirements by promoting robust model governance and risk management practices that include environmental considerations, thereby ensuring that banks maintain sustainable AI models aligned with regulatory standards.

The alignment of the RMF with these regulatory requirements also involves cross-departmental collaboration within the bank \cite{s38, s39, s40, s41}. To effectively include AI’s carbon footprint in the RMF, banks must foster collaboration between AI, sustainability, and risk management teams. AI and data science teams are responsible for developing and implementing energy-efficient AI models, tracking the carbon footprint, and integrating sustainability considerations into AI development practices. Sustainability teams oversee the bank's sustainability reporting, set carbon reduction targets for AI systems, and ensure alignment with CSRD, CSDDD, SS1/23, and the bank's ESG goals. Risk management teams are responsible for incorporating AI’s carbon footprint into the RMF, conducting regular risk assessments, and monitoring compliance with sustainability regulations.

Cross-departmental collaboration is facilitated by using advanced tools and technologies. For example, OxonFair \cite{s37} by Oxford Internet Institute\footnote{OxonFair: \url{https://github.com/oxfordinternetinstitute/oxonfair}} can be used for data governance, ensuring that data used in AI models is managed responsibly. Similarly, Fairlearn\footnote{FairLearn: \url{https://fairlearn.org/}} can be used to assess the fairness of AI systems, helping to identify and mitigate biases that may affect sustainability reporting. The GHG Protocol Toolkit can be used to measure the carbon emissions associated with AI models, providing the data needed for risk assessments and sustainability disclosures. Establishing a centralised dashboard to track AI-related carbon emissions, risk levels, and mitigation progress across departments can enhance collaboration and ensure a unified approach to managing the environmental risks of AI. Under the governance principles outlined in SS1/23, these tools support the development of a comprehensive risk management strategy that includes environmental sustainability as a core component \cite{s38, s39, s40, s41}.

\section{Conclusion}

Incorporating the carbon footprint of AI into the risk management frameworks of the banking sector is a critical step toward aligning AI deployment with sustainability objectives and regulatory requirements. The energy-intensive nature of AI systems poses significant environmental risks that must be recognised, measured, and managed within the bank's RMF \cite{s42, s43, s44}. By identifying, assessing, and mitigating the carbon footprint of AI, banks can ensure compliance with regulations such as the EU AI Act, CSRD, CSDDD, and SS1/23, while also mitigating reputational risks and promoting responsible AI use.

Effective incorporation of AI's carbon footprint into RMFs requires a multifaceted approach that includes risk identification, assessment, and mitigation strategies. It also requires integration with existing risk management processes, the establishment of monitoring and reporting mechanisms, and a clear definition of risk appetite and tolerance levels. Aligning these efforts with regulatory requirements ensures that banks not only comply with legal standards but also contribute to the broader goals of environmental sustainability and ethical AI deployment. Examples of the recent advances in AI, including more efficient computational frameworks like OLMoE, OneGen, and FSPAD, as well as smarter frameworks such as MemoRAG and agentic models like the Agentic RAG framework, present opportunities for banks to reduce the carbon footprint of their AI systems while enhancing performance.

Cross-departmental collaboration is essential for success in this endeavour. By fostering collaboration between AI, sustainability, and risk management teams, banks can develop and implement strategies that effectively manage the carbon footprint of AI, supporting both compliance and sustainability objectives \cite{s45, s46, s47}. The use of advanced tools and technologies further enhances the bank's ability to measure, monitor, and mitigate the environmental impact of AI systems. SS1/23’s emphasis on robust model governance, risk assessment, and lifecycle management further ensures that banks approach AI model management with sustainability in mind. By integrating the carbon footprint of AI into their RMFs, banks can take a proactive stance in addressing the environmental challenges of AI, ensuring that their use of AI supports a more sustainable and responsible future for the banking sector.

\bibliographystyle{alpha}
\bibliography{sample}

\end{document}